\documentclass{appolb}
\usepackage{epsfig}

\begin{document}
\title{ Evidence for Quark Gluon Plasma from Hadron Production
in High Energy Nuclear Collisions %
\thanks{Presented at Cracow School of Theoretical Physics,
May 29th -- June 8th, Zakopane, Poland }%
}
\author{ Marek Ga\'zdzicki
\address{ Institut f\"ur Kernphysik, University of Frankfurt \\
August Euler Str. 6, D--60486 Frankfurt, Germany }
}
\maketitle
\begin{abstract}
The  experimental results on the pion,
strangeness and $J/\psi$ production in high energy nuclear
collisions are discussed.
The anomalous energy dependence of pion and strangeness
production is consistent with the hypothesis that a transition 
to a deconfined phase takes place between the top AGS ($\approx$ 15 A$\cdot$GeV) and
the  SPS ($\approx$ 200 A$\cdot$GeV) energies.
The $J/\psi$ production systematics at the SPS can be understood assuming 
that the $J/\psi$ mesons are created at hadronization according
to the available hadronic phase space.
This new interpretation of the $J/\psi$ data allows one to
establish a coherent picture of high energy nuclear collisions
based on the statistical approaches of the 
collision early stage and hadronization.
Surprisingly,  the statistical model of strong interactions
is successful even in the region  reserved 
up to now for pQCD based models.

\end{abstract}
  
\section{Introduction}

The basic motivation for a broad experimental program  
of nucleus--nucleus (A+A) collisions at high energies
is the search for the Quark Gluon Plasma (QGP)
\cite{qgp}.
An impressive set of experimental data has been  collected 
during the last decades and many unexpected phenomena
have been discovered \cite{QM99}.
The  results indicate surprising scaling behaviours
which find a natural interpretation within statistical
models of the
early stage of the collision \cite{GaGo} as well as the hadronization 
\cite{Be:96,St:99}.
Within this framework one concludes that  the results are consistent
with the hypothesis of a QGP  creation in A+A collisions
at the SPS \cite{GaGo}.
The collision energy region in which the transition to QGP
takes place is located  between the top AGS ($\approx$ 15 A$\cdot$GeV) and 
the SPS ($\approx$ 200 A$\cdot$GeV)  energies.

This interpretation, however, is still under vivid discussion,
because the  statistical models
are not commonly recognized as  valid
tools to investigate high energy nuclear collisions.
Indeed, their basic assumptions cannot be derived 
from QCD.

On the other hand it is difficult to use QCD for the interpretation of the
experimental results. Problems arise because almost all effects
expected in the case of the transition to QGP are in the domain
of the so--called soft processes for which experimentally testable 
predictions of QCD are not available.
Attempts to build phenomenological QCD inspired models are not 
very successful \cite{Od:98} either.
Conclusive interpretation of the data within these models seems to be
impossible as one cannot estimate the uncertainties due to the used
approximations. 

Thus, the question whether QGP is created in A+A collisions at high energies
unavoidably leads  to the more fundamental question about our understanding
of strong interactions.

The aim of this contribution is a brief discussion 
within the framework of the statistical models
of the data on the pion,
strangeness and $J/\psi$ production in nuclear collisions.

\section{Pion Production}

The majority of particles  produced during high energy interactions
are pions.
Thus, pions carry basic information on entropy created in the collision
and consequently their yield should be sensitive to the effective number
of degrees of freedom at the early stage. 
The energy dependence of mean pion multiplicity in nucleon--nucleon (N+N)
interactions is plotted in  Fig. 1 \cite{GaRo}. 
The pion yield appears to be  proportional the energy measure introduced
by Fermi \cite{Fe:50}:
\begin{equation}
F \equiv (\sqrt{s}_{NN} - 2 m_N)^{3/4}/\sqrt{s}_{NN}^{1/4},
\end{equation}
where $\sqrt{s}_{NN}$ is the c.m. energy for a nucleon--nucleon pair and
$m_N$ is the nucleon mass.
This dependence was predicted by Fermi \cite{Fe:50} 
and Landau  \cite{La:53}  almost 50 years ago.
It follows directly from the assumption that the most probable
(maximum entropy) state is created in the early stage of the collision.
The energy dependence of the pion yield in central A+A collisions
is different from that observed for N+N interactions.
The comparison is presented in Fig. 2 \cite{GaRo} where the difference
between the average number of pions per wounded nucleon (participant)
in A+A and N+N interactions
\cite{Bi:76}
is shown as a function of the collision energy.
At low  energies (the AGS and below) the pion production in 
A+A collisions is significantly suppressed in comparison to N+N interactions.
This suppression can be understood 
as due to  entropy transfer to the baryonic sector \cite{GGM}
during the expansion of the matter \cite{Le:90}. 
The pion enhancement effect is observed in central A+A collisions
at the SPS \cite{add4}.
The change from the pion suppression to pion enhancement pattern
can be attributed to the transition 
to the QGP occurring between the AGS and the SPS energies.
In fact, the statistical model of the early stage, which assumes
this transition, correctly reproduces the data \cite{GaGo}, see solid line in Fig. 2.
In the model the pion enhancement is due to the increased entropy content
of the deconfined matter.

\section{Strangeness Production}

The energy dependence of the strangeness to pion ratio in A+A
and N+N interactions is compared in Fig. 3 \cite{GaRo} where the
ratio
\begin{equation}
E_S \equiv
( \langle \Lambda \rangle + \langle K
+ \overline{K} \rangle )/\langle \pi \rangle
\end{equation}
is plotted as a function of $F$.
Between the AGS 
($F \approx 2$) and the SPS ($F \approx 4$)
energies the ratio for N+N interactions increases
by a factor of about 2.
Very different behaviour is observed for central A+A collisions,
where no significant energy dependence is observed.
The latter behaviour can be interpreted as due to the transition
to the QGP taking place between the AGS and the SPS  energies.
The statistical model of the early stage suggests that the
ratio should reach a maximum at the beginning of the transition region \cite{GaGo},
see solid line in Fig. 3.
Thus, a  non--monotonic energy behaviour of the strangeness
to pion ratio is predicted;
the transition is signal by the suppression of the strangeness
to pion ratio.

We note that in the picture of the non--equilibrium strangeness
production a very different conclusion is reached \cite{Ra:86},
namely
the strangeness enhancement is expected to be a signal of the
transition.

\section{$J/\psi$ Production}

Since a long time the production of $J/\psi$ mesons has been
considered
as a sensitive probe of the state of matter in the early stage \cite{Sa:86, Sh:78}.
This reasoning is based on the assumption that the $J/\psi$ meson 
(or its pre-state) is produced at the very beginning of the
collision process due to the coalescence of the $c$--$\bar{c}$
pairs produced in the hard QCD process \cite{Ma:95}.
This orthodox picture suggests also that the Drell--Yan pairs
should be used as a proper  reference for the $J/\psi$
study.
Within this framework a complicated pattern of $J/\psi$
suppression was established experimentally by the NA38 and NA50 Collaborations
\cite{NA38, NA50}.
The analysis of this pattern  suggests that the deconfined matter is created 
only in central Pb+Pb collisions where an 
anomalous suppression was observed.

This conclusion as well as the theoretical 
framework leading to it
are very different from 
the pion and strangeness cases discussed in the previous
sections.
Thus, the  picture of the A+A collision process seems to be
inconsistent.

It has been, however, 
 recently found \cite{GaGo, Ga} that the $J/\psi$ multiplicity increases
proportionaly to the pion multiplicity in collisions from p+p
to central Pb+Pb at the  SPS.
The ratio of the $J/\psi$ multiplicity to the multiplicity of
negatively charged hadrons (mostly $\pi^-$ mesons) is plotted
in Fig. 4 \cite{GoGa}.
The approximate independence of the ratio of must be 
accidental when considered in the orthodox picture
of $J/\psi$ production.

This scaling behaviour
finds, however, a natural explanation in the model of the statistical
production of $J/\psi$ mesons at the hadronization \cite{GoGa}.
We recall here that the statistical models of hadron production
(hadronization) are successfully used to describe the data
from the central Pb+Pb collisions as well as from the interactions of elementary
particles (p+p, e$^+$+e$^-$).
The  temperature parameter which defines the available
phase--space is found to be energy and system size independent
(for sufficiently high energies) \cite{Be:96},
it ranges between 160--190 MeV.
In order to reproduce the measured $J/\psi$ yield the temperature parameter
$T \approx 176$ MeV is needed \cite{GoGa}. 
It is in good agreement with the values obtained
in the analysis of hadron yield systematics.
The statistical approach to the $J/\psi$ creation  allows for
a coherent interpretation of the results on pion, strangeness and
$J/\psi$ production in nuclear collisions.
It changes however in a significant way the role of the $J/\psi$ meson.
As being produced directly at the hadronization it is not sensitive
to the form of prehadronic matter.
But due to its large mass and small cross section for hadronic interactions
it can serve as a sensitive probe of the hadronization process.

\section{Event--by--Event Fluctuations}

The question whether statistical models can serve as a valid description
of the A+A collision process is crucial in the interpretation of
the experimental results.
The discussion presented above was based on 
the  analysis of the inclusive data on hadron production,
i.e. the  results obtained by averaging over a class of selected events
(e.g. central collisions).
A key test of the validity of the statistical
approach can be done by the study  of event--by--event fluctuations.
Large acceptance, high statistics and high quality data obtained by the NA49
experiment \cite{nim} allow one to perform such an  analysis of many observables.

The event--by--event fluctuation of the mean transverse
momentum and the kaon to pion ratio for central Pb+Pb collisions
at 158 A$\cdot$GeV are shown in Fig. 5 \cite{na49pl, cr}.
The data are compared with the fluctuations simulated for
the case of  independent
particle production where the multiplicity distribution of generated
events is equal to the measured one (the 'mixed event' procedure).
This  procedure gives  fluctuations as expected in the statistical
model within the grand canonical ensemble \cite{Mr:98}  when the small effects
due to quantum statistics, Coulomb interaction and resonance decays
are neglected.
The measured fluctuations indeed appear to be close to those
expected in the statistical models, see Fig. 5.

Systematic, quantitative study of event--by--event fluctuations
is  done using the $\Phi$ measure of fluctuations \cite{GaMr}.
It allows to remove the influence of 'unwanted' fluctuations
of the volume (number of wounded nucleons) of colliding nuclei.
The values of $\Phi_{P_T}$ (the transverse momentum
fluctuation measure) obtained for all inelastic p+p interactions
and central Pb+Pb collisions at 158 A$\cdot$GeV are shown in Fig. 6 as
a function of the number of wounded nucleons \cite{na49pl}.
The influence of short range correlations due to quantum statistics
and Coulomb interaction is removed from the result.
The value of $\Phi_{P_T}$  for central Pb+Pb collisions is consistent
with zero, the value of $\Phi_{P_T}$ calculated in the statistical model
in grand canonical ensemble for classical particles \cite{Mr:98}.
The non-zero, positive value of $\Phi_{P_T}$  for p+p interactions
can be attributed to the influence of the energy--momentum conservation,
its role appears to be significant for the small systems \cite{liu}.
Calculations in the micro-canonical ensemble
are needed here.
In the approaches where the A+A collisions are modelled as an independent
superposition of N+N interactions the value of $\Phi$ 
in A+A collisions is equal to the corresponding value for the
elementary process \cite{GaMr}.

We conclude that the fluctuations measured in central Pb+Pb collisions
at the SPS confirm the validity of the statistical approach used to
interpret the inclusive results.

\section{Summary}

Statistical models of the early stage and the hadronization in high energy
nuclear collisions allow one  a coherent interpretation
of the wide spectrum of the experimental data.
Within this interpretation:
\begin{itemize}
\item
the energy dependence of pion and strangeness yields serves as 
an evidence for  Quark Gluon Plasma creation in A+A collisions at
the SPS;
\item
the transition to the deconfined state takes place between the top AGS and
the SPS energies and  should be reflected by a non--monotonic 
dependence of the strangeness to pion ratio in the intermidiate
energy region;
\item
the systematics of the $J/\psi$ production can be understood 
assuming the statistical creation of the $J/\psi$ mesons at the hadronization;
within this interpretation the yield of $J/\psi$ mesons
is independent of the properties of the early stage matter,
but it is sensitive to  the hadronization
process.

\end{itemize}

\vspace{0.2cm}

Analysis of the event--by--event fluctuations in high energy collisions
yields a crucial, independent test  of the validity of the
statistical approach.
The observed fluctuations are consistent with those expected
in the statistical models.

\vspace{0.2cm}

Finally, we repeat once more that the interpretation discussed
above is based on the analysis  within statistical models.
The validity of this approach is however controversial.
Thus, the question whether QGP is created in A+A collisions at the  SPS
leads us to questions about our understanding of strong interactions.
It is clear that the increasing flow of  new experimental data 
on nuclear (p+p, p+A and A+A)  collisions should
soon result in a substantial progress in this domain.

\vspace{0.5cm}
I would like to thank organizers of this school for a very
intresting and stimulating meeting.
I thank St. Mr\'owczy\'nski and P. Seyboth for comments.

\newpage
 
\begin{figure}[p]
\epsfig{file=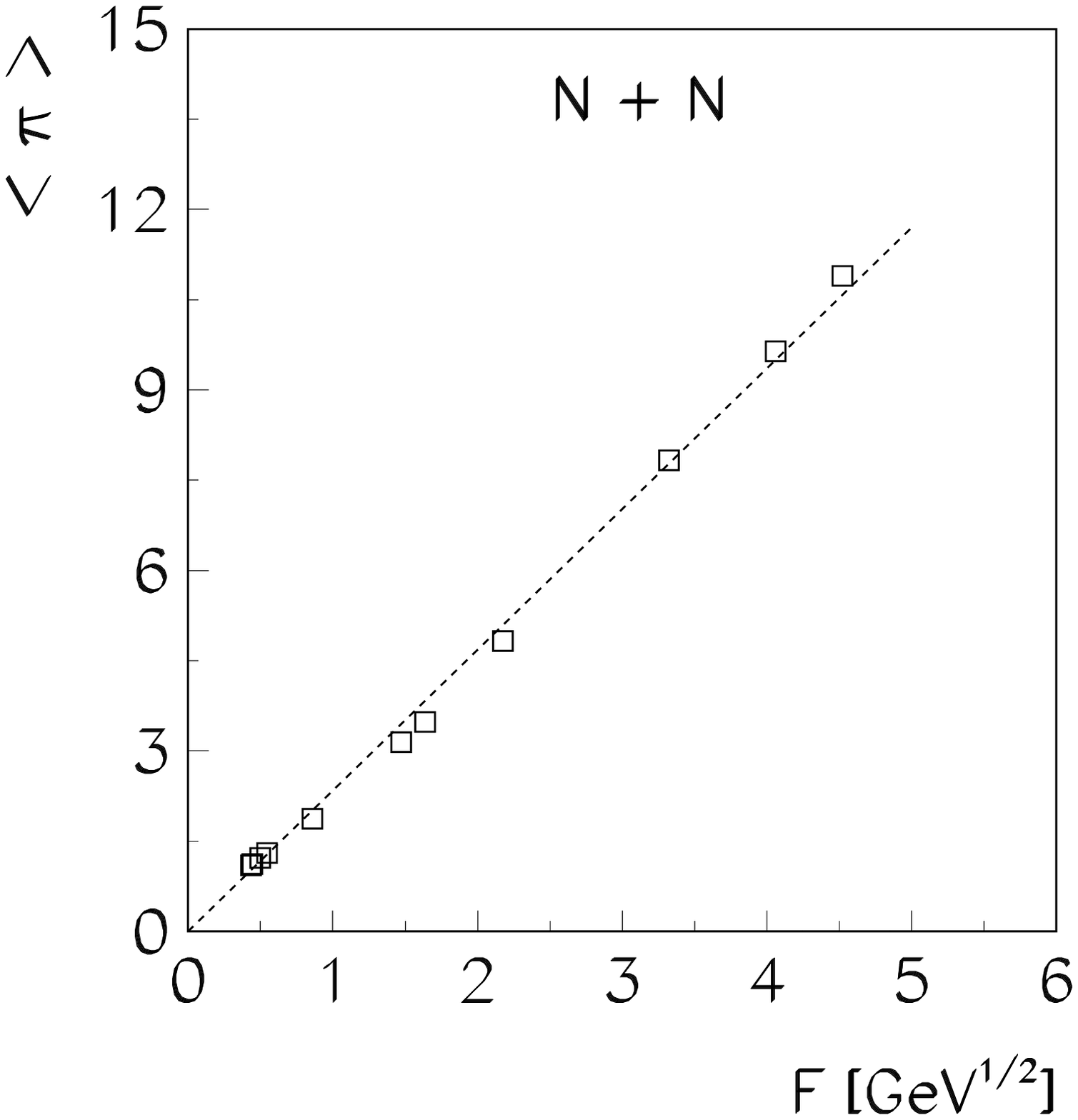,width=12cm}
\caption{
The dependence of the mean pion multiplicity
for all inealstic nucleon--nucleon interactions on
the collision energy  measured
by the Fermi energy variable,
$F = (\sqrt{s}_{NN} - 2 m_N)^{3/4})/ \sqrt{s}_{NN}^{1/4} $.
The dashed line is plotted to guide the eye.
 }
\label{fig1}
\end{figure}

\newpage

\begin{figure}[p]
\epsfig{file=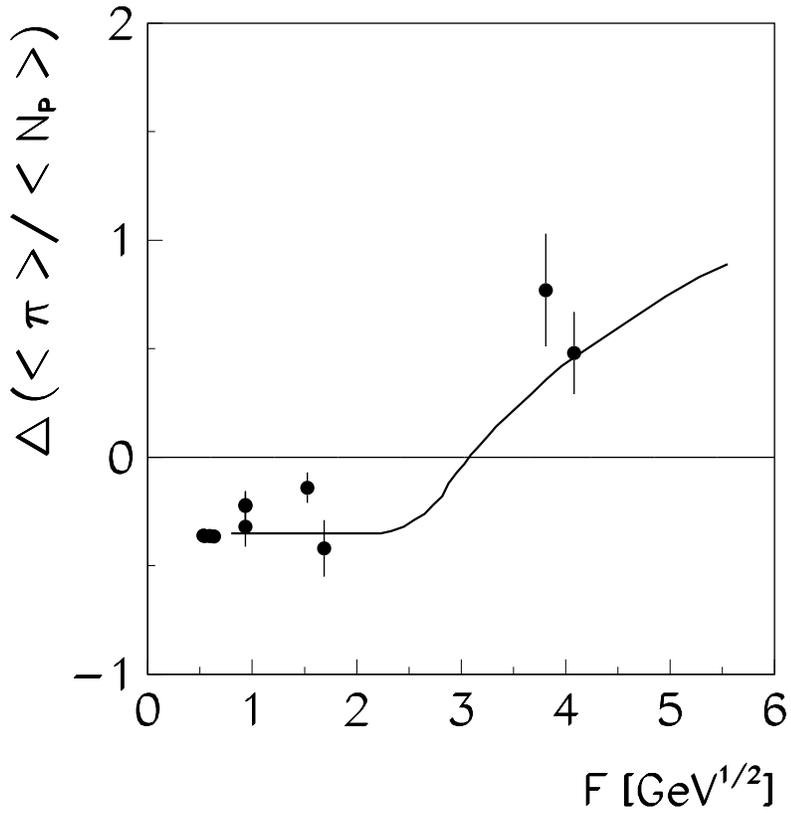,width=12cm}
\caption{
The dependence of the difference
between pion/baryon ratios for central A+A collisions
and nucleon--nucleon interaction at the same energy per nucleon  on the
collision energy  \protect\cite{GaRo} measured
by the Fermi energy variable,
$F = (\sqrt{s}_{NN} - 2 m_N)^{3/4})/ \sqrt{s}_{NN}^{1/4} $.
The solid line shows predictions of the statistical
model of the early stage assuming transition to the QGP
between the top AGS ($F \approx$ 2) and SPS ($F \approx$ 4).
 }
\label{fig2}
\end{figure}

\newpage

\begin{figure}[p]
\epsfig{file=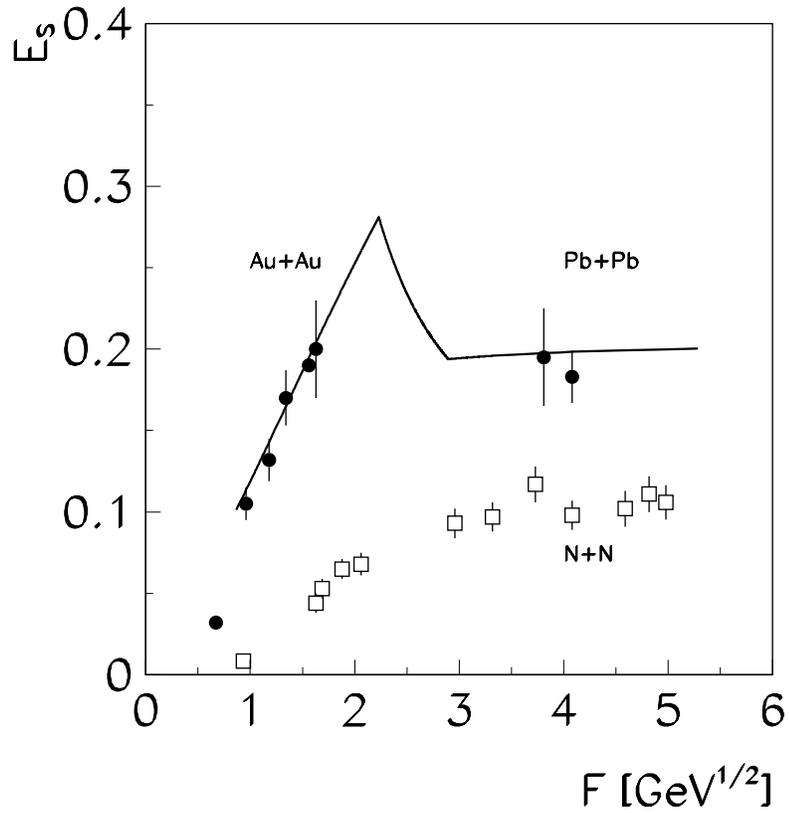,width=12cm}
\caption{
The dependence of the
strangeness/pion ratio,
$E_S = (\langle \Lambda \rangle + \langle K + \overline{K} \rangle) /
\langle \pi \rangle$,
for central A+A collisions (closed circles)
and nucleon--nucleon interactions (open squares) as a function of
collision energy  \protect\cite{GaRo} measured by the Fermi energy variable,
$F$.
The solid line shows predictions of the statistical
model of the early stage assuming transition to the QGP
between the top AGS ($F \approx$ 2) and SPS ($F \approx$ 4).
 }
\label{fig3}
\end{figure}

\newpage

\begin{figure}[p]
\epsfig{file=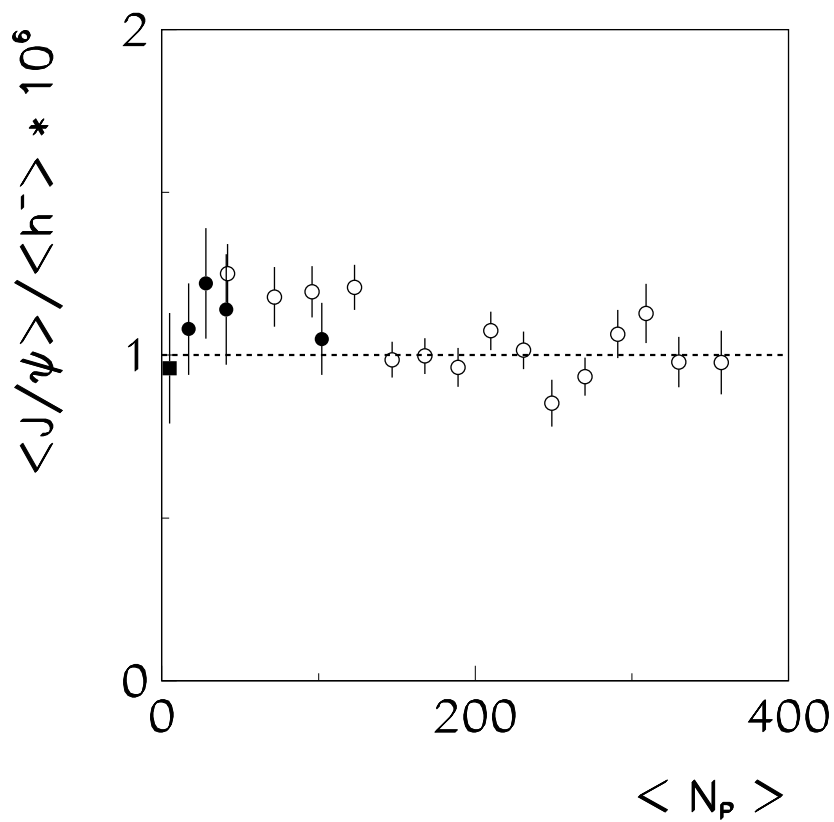,width=12cm}
\caption{
The ratio of the mean multiplicities of $J/\psi$ mesons
and negatively charged hadrons for inelastic nucleon--nucleon (square) and
inelastic O+Cu, O+U, S+U and Pb+Pb (circles) interactions at
158 A$\cdot$GeV plotted as a function of the mean
number of participant nucleons.
For clarity the N+N point is shifted from
$\langle N_P \rangle = 2$ to $\langle N_P \rangle = 5$.
The dashed line indicates the mean value of the ratio.
}
\label{fig4}
\end{figure}

\newpage

\begin{figure}[p]
\epsfig{file=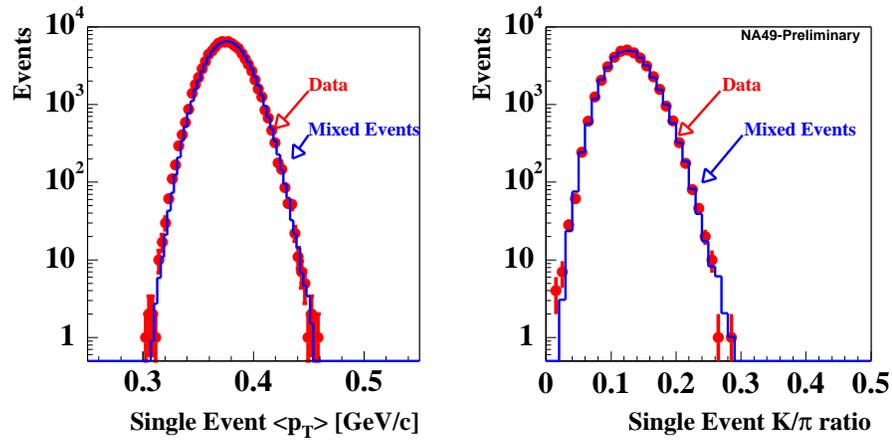,width=12cm}
\caption{
The event--by--event fluctuations of the mean transverse
momentum and the kaon to pion ratio for central Pb+Pb collisions 
at 158 A$\cdot$GeV.
The solid lines indicate fluctuations calculated assuming independent
particle emission (the 'mixed event' procedure).
 }
\label{fig5}
\end{figure}

\newpage

\begin{figure}[p]
\epsfig{file=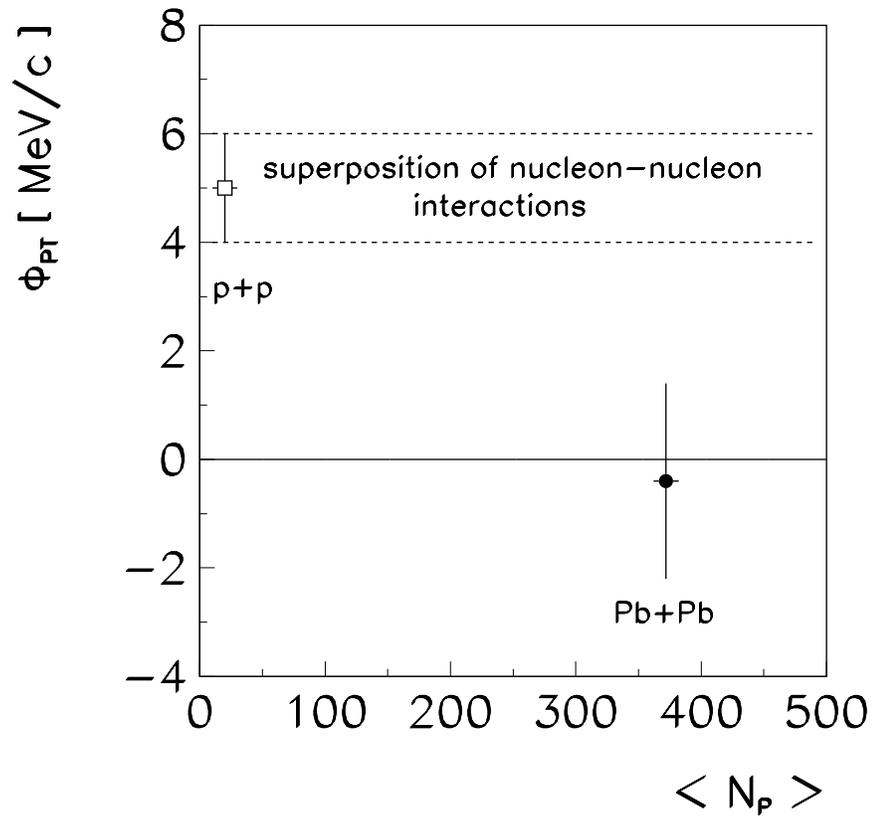,width=12cm}
\caption{
The $\Phi_{p_T}$ fluctuation measure dependence on the number
of wounded nucleons. The two data points show results of the
NA49 Collaboration for all inelastic p+p interactions and
central Pb+Pb collisions at  158 A$\cdot$GeV.
 }
\label{fig6}
\end{figure}

\end{document}